\documentclass[%
  prl,
 amsmath,amssymb,
 reprint,%
]{revtex4-1}
\usepackage{graphicx}
\usepackage{dcolumn}
\usepackage{bm}
\usepackage{comment}
\usepackage{color}
\usepackage{float}
\usepackage{soul}
\usepackage[utf8]{inputenc}
\usepackage[T1]{fontenc}
\usepackage{mathptmx}
\usepackage{subfig}
\usepackage[inline]{enumitem}
\usepackage{caption}
\newcommand{\bogus}[1]{{}}

\begin{document}


\title{Collisionless cooling of perpendicular electron temperature in the thermal quench of a magnetized plasma}

\author{Yanzeng Zhang}%
\affiliation{Theoretical Division, Los Alamos National Laboratory, Los Alamos, New Mexico 87545, USA}
\author{Jun Li}%
\affiliation{Theoretical Division, Los Alamos National Laboratory, Los Alamos, New Mexico 87545, USA}
\affiliation{School of Nuclear Science and Technology, University of Science and Technology of China, Hefei, Anhui, China~\footnote{Present address}}
\author{Xian-Zhu Tang}%
\affiliation{Theoretical Division, Los Alamos National Laboratory, Los Alamos, New Mexico 87545, USA}


\begin{abstract}
  Thermal quench of a nearly collisionless plasma against a cooling
  boundary or region is an undesirable off-normal event in magnetic
  fusion experiments, but an ubiquitous process of cosmological
  importance in astrophysical plasmas.  There is a well-known mismatch
  that what experimentally diagnosed is the drop in perpendicular
  electron temperature $T_{e\perp},$ but the parallel transport theory
  of ambipolar-constrained tail electron loss produces parallel
  electron temperature $T_{e\parallel}$ cooling.  Here two
  collisionless mechanisms, dilutional cooling by infalling cold
  electrons and wave-particle interaction by two families of whistler
  instabilities, are shown to enable fast $T_{e\perp}$ cooling that
  closely tracks the mostly collisionless crash of $T_{e\parallel}.$
\end{abstract}

\maketitle

Magnetic confinement of a fusion-grade plasma in the Laboratory has
shown that extreme care must be exercised in the design of the
magnetic fields in order to sustain a nearly collisionless
plasma~\cite{boozer-rmp-2004,Helander_RPP_2014}. In space and
astrophysical systems, the low particle density and extremely large
spatial scale can easily accommodate nearly collisionless
plasmas~\cite{Parks-ency-atom-science-2015,denton-etal-JGRSP-2010,denton-cayton-AG-2011,borovsky-etal-JGRSP-2016,fabian-springer-2002,Peterson-Fabian-PR-2006,Fabian-ARAA-1994,Hitomi-collaboration-nature-2016,Zhuravleva-etal-nature-2014}.
Large scale cooling of such nearly collisionless plasmas, especially
in the presence of structure formations in a tenuous astrophysical
plasma background, becomes a plasma transport process of significant
cosmological
importance~\cite{Fabian-ARAA-1994,Peterson-Fabian-PR-2006,Hitomi-collaboration-nature-2016,Zhuravleva-etal-nature-2014}.
One of the most interesting features is the so-called cooling flow,
the presence of which defies the normal transport
closure~\cite{Binney-Cowie-apj-1981,Fabian-etal-AAR-1991,Fabian-ARAA-1994,Peterson-Fabian-PR-2006}
such as the collisional Braginskii~\cite{braginskii} and collisionless
flux-limiting~\cite{atzeni_book_2004} forms of electron thermal
conduction, but is allowed if the plasma kinetics is fully accounted
for~\cite{Zhangfronts}. Interestingly, a wholly undesirable phenomenon
in the magnetic confinement experiment, the so-called
thermal quench (TQ) in the first phase of a tokamak
disruption~\cite{hender2007mhd,nedospasov2008thermal}, provides a
laboratory platform to understand the intriguing {\em plasma kinetics} underlying the
rapid cooling of a nearly collisionless plasma against a cooling
boundary, which can be the chamber wall or injected high-Z
pellets~\cite{federici2003key,baylor2009pellet}.

The millisecond and sub-millisecond time-scale TQ~\cite{shimada2007chapter,riccardo2005timescale,paz2020runaway} of
a magnetically confined plasma is thought to be dominated by plasma
parallel transport, especially electron thermal conduction, along open
(stochastic) magnetic field lines. The most extreme and
astrophysically relevant regime has the magnetic connection length
$L_B$ comparable to or even significantly shorter than the core plasma
mean-free-path $\lambda_{mfp}.$ The conventional wisdom is that in
such low collisionality regime, the electron parallel conduction flux
would follow the so-called flux-limiting (FL) form, $q_{en} \sim n_e
T_{\parallel e} v_{th,e\parallel}, $ with $v_{th,e\parallel}=\sqrt{
  k_B T_{e\parallel}/m_e}$ the local  parallel electron thermal
speed~\cite{atzeni_book_2004,Bell-pof-1985}. As a result, the TQ time
would follow the scaling $\tau_{TQ}^{FL} \propto m_e^{1/2}
(n_0\ln\Lambda)^{-1/4}L_B^{3/4}T_0^{0}$ with $n_0$ and $T_0$ the
initial plasma density and temperature, respectively, and $\ln
\Lambda$ the Coulomb logarithm, but the cooling flow can not be supported.
Recent simulations and
analysis~\cite{Zhangfronts,listaged} showed instead that ambipolar
transport constrain the electron parallel thermal conduction so that
the cooling flow is supported, and $\tau_{TQ}^\parallel \sim
\left(K_{n0}\sqrt{m_i/m_e}\right)^{1/4} L_B/c_s$ with $c_s$ the
initial ion sound speed. The initial Knudsen number
$K_{n0}=\lambda_{mfp}/L_B$ with $\lambda_{mfp} \propto T_0^2/(n_0\ln
\Lambda)$ so $\tau_{TQ}^\parallel \propto m_i^{3/4}m_e^{-1/4}\left(n_0
\ln\Lambda \right)^{-1/4} L_B^{3/4} T_0^0.$ Remarkably,
a recent analysis of EAST disruption experiments~\cite{xia2023timescale} revealed
an extremely weak dependence  of $\tau_{TQ}$ on $T_0$, $\tau_{TQ}^{\perp, exp} \propto
T_0^{-0.08},$ consistent with the predicted scaling with $T_0$
($\tau_{TQ}^\parallel\propto T_0^0$), notwithstanding that the critical $m_i/m_e$ scaling
in $\tau_{TQ}^\parallel$ in relation to $\tau_{TQ}^{FL},$ as well as the presence/absence of a cooling flow,
remain to be checked by experiments.

The apparent agreement in $T_e$ scaling of $\tau_{TQ}$ actually belies
a critical physics gap between theory and experimental measurements,
in that parallel thermal conduction in a nearly collisionless plasma
cools
$T_{e\parallel}$~\cite{CGL-prsl-1956,tang-ppcf-2011,guo2012ambipolar,Zhangfronts,listaged}
but the electron cyclotron emission (ECE)
diagnostics~\cite{riccardo2005timescale,paz2020runaway,xia2023timescale}
measure $T_{e\perp}.$ {\em Collisional} cooling of $T_{e\perp}$ can be
fast once $T_{e\parallel}$ becomes sufficiently low, since collisional
electron temperature isotropization follows $\partial
T_{e\perp}/\partial t = - \left(T_{e\perp} -
T_{e\parallel}\right)/\tau_{e\perp}^c$ with $\tau_{e\perp}^c \propto
m_e^{1/2}\left(n_e \ln\Lambda\right)^{-1}
T_{e\parallel}^{3/2}$~\cite{Chodura_1971,Li-etal-prl-2022}.  Indeed,
the collisional cooling ($\tau_{e\perp}^c$) of $T_{e\perp}$ would be
mostly independent of the initial core temperature $T_0$ and can be
quite fast if $T_{e\parallel}$ becomes sufficiently low. In this
scenario, the TQ is separated into two distinct phases,
with durations $\tau_{TQ}^\parallel$ and $\tau_{TQ}^\perp=\tau_{e\perp}^c,$ both of
which have no or weak dependence on $T_0,$ but with very
different scalings with respect to $L_B$ ($\tau_{TQ}^\parallel \propto
L_B^{3/4}$ v.s.  $\tau_{TQ}^\perp\propto L_B^0$) and plasma density
($\tau_{TQ}^\parallel\propto n_0^{-1/4}$ v.s. $\tau_{TQ}^\perp \propto
n_0^{-1}$). How short $\tau_{TQ}^\perp$ can be is mostly set by how
low $T_{e\parallel}$ can be cooled in the first phase.

This Letter describes the physics of {\em collisionless} cooling of
$T_{e\perp},$ which produces qualitatively different TQ history in
that there is no longer a separate phase for $T_{e\perp}$ as
$T_{e\parallel,\perp}$ now follow the same $\tau_{TQ}$ scaling
previously given for $T_{e\parallel},$ {\em i.e.} $\tau_{TQ}\approx
\tau_{TQ}^\parallel \propto m_i^{3/4} m_e^{-1/4} n_0^{-1/4} L_B^{3/4}
T_0^0$.  Our analysis indicates that this is an unavoidable scenario
as long as $L_B \le \lambda_{mfp}$ at the onset of TQ.  There are
actually two distinct mechanisms that can contribute to collisionless
cooling of $T_{e\perp}.$ The first is the infalling cold electrons
from the cooling zone which has a much lower temperature $T_w \ll
T_0$. They follow the ambipolar electric field into the core plasma
and reduce the core $T_{e\perp}$ by dilutional cooling.  The second
mechanism is the result of electron temperature isotropization via
wave-particle interaction, by self-excited electromagnetic waves in
the whistler range. There are actually two distinct types of whistler
instabilities involved. What gets excited first is the
trapped-electron whistler (TEW) mode, previously identified
in~\cite{guo2012ambipolar}, but is now significantly modified by the
infalling cold electrons.  This first type of whistler instabilities
drives the truncated electron distribution towards a bi-Maxwellian
with $T_{e\perp}\gg T_{e\parallel}.$ The ensuing, second type of
whistler instability is the well-known temperature anisotropy driven
whistler (TAW) mode~\cite{kennel1966limit,gary1996whistler}, which can
aggressively bring down $T_{e\perp}.$ Upon the nonlinear saturation of
the whistler instabilities, $T_{e\perp}/T_{e\parallel}$ approaches the
marginality of TAW, which depends on the plasma beta. As a general
guidance, for modest cooling ($T_0/T_w \lesssim 10$), dilutional
cooling is sufficient to align $T_{e\perp}$ cooling with that of
$T_{e\parallel}.$ Deep cooling ($T_0/T_w \gtrsim 10^2$) critically
relies on the two types of whistler instabilities to work in sequence.

To elucidate the physics of collisionless $T_{e\perp}$ cooling, we
first briefly review how $T_{e\parallel}$ is rapidly cooled by
parallel transport.  In a nearly collisionless plasma,
$T_{e\parallel}$ cooling is the result of tail electron loss~\cite{Zhangfronts}, which
produces a truncated distribution function in $v_\parallel$ that has
the cutoff speed $v_c = \sqrt{2e \Delta\Phi_{RF}/m_e}$ so
$f_e(v_\parallel > v_c)=0.$ Here the reflecting potential
$\Delta\Phi_{RF}$ arises in order to enforce ambipolar transport, and
a decreasing $\Delta\Phi_{RF}$ leads to $T_{e\parallel}$ cooling. For
deep cooling, i.e. $T_{e\parallel} \ll T_0,$ the reflecting potential
satisfies $v_c \ll v_{th,e}\equiv\sqrt{k_BT_0/m_e}.$
In the middle of the open magnetic field line of connection length $L_B,$
the electrostatically trapped electron distribution of the core plasma can thus be modeled as
\begin{align}
f_t(v_\parallel,v_\perp)=\frac{2
  n_e}{\textup{erf}(v_c/v_t)\sqrt{\pi}v_t^3}e^{-(v_\parallel^2+v_\perp^2)/v_t^2}\Theta(1-v_\parallel^2/v_c^2),\label{eq_cutoff_distribution}
\end{align}
where $v_t=\sqrt{2}v_{th,e}, ~n_e$ is the electron density, and
$\textup{erf}(x)$ and $\Theta(x)$ are the error and Heaviside step
function, respectively.

{\em $T_{e\perp}$ cooling by dilution:}
In a nearly collisionless plasma, the cold electrons near the cooling
boundary, where electron energy is taken out by impurity
radiation and/or wall recycling, will move upstream into the core
plasma by following the ambipolar electric field, gaining the kinetic
energy of $e \Delta \Phi_{RF}.$ These infalling cold electrons
can be modeled as
\begin{equation}
f_r^\pm=\frac{2
n_e}{v_{w}^2v_c}e^{-v_\perp^2/v_{w}^2}\delta(1\pm
v_\parallel/v_c),\label{eq_reflected_distribution}
\end{equation}
where $v_{w}=\sqrt{2T_w/m_e}$ 
and
$\delta(x)$ is the delta-function. Noting that
$\int_{-\infty}^\infty\int_0^\infty \left(f_t, f_r^\pm\right) v_\perp dv_\perp dv_\parallel
=n_e,$  we can parameterize
the fraction of the cold electron beam density by $\alpha$ so that the
total core electron distribution is
\begin{align}
  f_e=(1-\alpha) f_t + \alpha (f_r^++f_r^-)/2. \label{eq:fe-sum}
\end{align}
During TQ, one can assume that $f_e$ at $v_\parallel=0$ doesn't
change~\cite{Zhangfronts}, so $n_e/\textup{erf}(v_c/v_t)\times
(1-\alpha)=n_0$. This implies
$\alpha\leq \alpha_{max}=1-\textup{erf}(v_c/v_t)$ since $n_e\leq n_0$.
One then finds that the infalling cold
electrons would dilutionally cool the core $T_{e\perp}$ to $\alpha
T_{w} + (1-\alpha) T_0,$
with the constraint of $\alpha \le \alpha_{max}$ as noted earlier.

{\em Modification of trapped electron whistler (TEW) instability by
  infalling cold electron beams:} The truncated electron distribution
$f_t$ of Eq.~(\ref{eq_cutoff_distribution}) is known to drive robust whistler
instabilities~\cite{guo2012ambipolar}. To understand the impact of the infalling cold
electron population $f_r^\pm$ of
Eq.~(\ref{eq_reflected_distribution}), we substitute $f_e$ of
Eq.~(\ref{eq:fe-sum}) into the dispersion of whistler wave along the magnetic
field with normal mode $\exp\left(ik x- i\omega t\right),$~\cite{Krallpriciples}
\begin{align}
&0=1-\frac{k^2c^2}{\omega^2}+\\\nonumber
&\frac{\omega_{pe}^2}{n_e\omega}\int\int\left[\left(1-\frac{kv_\parallel}{\omega}\right)\frac{\partial
    f_e}{\partial v_\perp^2}+\frac{kv_\parallel}{\omega}\frac{\partial
    f_e}{\partial
    v_\parallel^2}\right]\frac{v_\perp^3 dv_\perp dv_\parallel}{\omega-kv_\parallel
  -\omega_{ce}},\label{eq-dispersion-def}
\end{align}
where we have ignored the effect of ions assuming $\omega_{ci}\ll
\omega<\omega_{ce}$ with $\omega_{ce,i}$ the electron/ion
gyro-frequency, and $\omega_{pe}$ is the plasma frequency. This leads to the dispersion relation
\begin{equation}
D(\omega,k)=1-\frac{k^2c^2}{\omega^2}+(1-\alpha)D_t+\alpha
D_r=0,\label{eq-dispersion-trap-reflect}
\end{equation}
where
\begin{align}
D_t&=\frac{\omega_{pe}^2}{\textup{erf}(\hat{v}_c)\sqrt{\pi}\omega^2}\left[\frac{\omega}{kv_t}\int_{-\hat{v}_c}^{\hat{v}_c}\frac{e^{-\hat{v}_\parallel^2}}{\hat{v}_\parallel-\xi}d\hat{v}_\parallel
  +\frac{\hat{v}_ce^{-\hat{v}_c^2}}{\hat{v}_c^2-\xi^2}\right],\label{eq-dispersion-trap}\\
D_r&=-\frac{\omega_{pe}^2}{\omega^2}\left[\frac{\hat{v}_c^2-\omega
    \xi/(kv_t)}{(\hat{v}_c+\xi)(\hat{v}_c-\xi)}
  +\frac{(\xi^2+\hat{v}_c^2)\hat{v}_{w}^2}{2[(\hat{v}_c+\xi)(\hat{v}_c-\xi)]^2}
  \right],\label{eq-dispersion-reflec-delta-func}
\end{align}
with $\hat{v}_{\parallel, c, w}=v_{\parallel, c,w}/v_t$, and
$\xi=(\omega-\omega_{ce})/kv_t$. 

The contribution of the infalling cold electron beams can be examined by setting $\alpha=1.$
Important insights can be readily obtained in the limiting cases of 
 $\hat{v}_c\ll 1$ and $\hat{v}_c\gg 1$. For the former case, both
$\xi+ \hat{v}_c$ and $\xi-\hat{ v}_c$ approximate to
$i\gamma/(kv_t)$ and thus contribute equally to $D_r$, where $i\gamma/(kv_t)>\hat{v}_c$ with $\xi\approx
0$. Let's further
assume that $|\omega\xi/(kv_t)|\ll \hat{v}_{w}^2$, in the limit of
$k^2c^2\gg \omega^2$ we obtain
\begin{equation}
\gamma=\frac{v_{w}}{\sqrt{2} c} \omega_{pe}.\label{eq-growth-reflec-only-small-vc}
\end{equation}
Notice that the growth rate in
Eq.~(\ref{eq-growth-reflec-only-small-vc}) is similar to that of TEW~\cite{guo2012ambipolar} with
$v_{w}$ replacing $v_{t}$ as the free energy for the instability. This is not surprising since both $f_t$ and
$f_r^{\pm}$ are like delta-function in $v_\parallel$ for whistler
modes in the limit of $v_c\ll \omega/k\sim v_t$. On the other hand, for large $v_c $,
only one of $\hat{v}_c\pm\xi$ satisfies the resonant condition, so
\begin{equation}
\gamma=\frac{v_{w}}{2c}\omega_{pe},\label{eq-growth-reflec-only-large-vc}
\end{equation}
for $k^2c^2\gg\omega_r^2$. The different factor in
Eqs.~(\ref{eq-growth-reflec-only-small-vc},
\ref{eq-growth-reflec-only-large-vc}) results from the different
number of resonant conditions. In contrast to the TEW instability~\cite{guo2012ambipolar} for $v_c\gg v_t$,
where the growth rate $\gamma\propto \exp(-\hat{v}_c^2/2)$
significantly decreases with increasing $\hat{v}_c$, $\gamma$ in
Eq.~(\ref{eq-growth-reflec-only-large-vc}) is independent of
$\hat{v}_c$. 
 In
reality, the decompressional cooling of $T_{e\parallel}^{beam}$ for
the infalling cold electrons will produce a lower
$T_{e\parallel}^{beam}<T_w$ compared with the $T_{e\parallel}^{beam}=T_w,$ but
not $T_{e\parallel}^{beam}=0$ as represented by $\delta(1\pm
v_\parallel/v_c)$ in Eq.~(\ref{eq_reflected_distribution}). The
$\gamma$s in
Eqs.~(\ref{eq-growth-reflec-only-small-vc},\ref{eq-growth-reflec-only-large-vc})
are thus upper bounds for a quantitative estimate.

Comparing Eqs.~(\ref{eq-growth-reflec-only-small-vc},
\ref{eq-growth-reflec-only-large-vc}) with the growth rates of the
pure TEW mode ($\alpha=0$) in~\cite{guo2012ambipolar}, we find that the impact of the
infalling cold electrons for small but finite $\alpha$ on TEW instability depends on
$\hat{v}_c$. For small $\hat{v}_c\ll 1$, we have
\begin{align}
\gamma= R \frac{v_t}{\sqrt{2} c}
\omega_{pe},\label{eq-growth-reflect-trap-small-vc}
\end{align}
with $R\equiv \sqrt{(1-\alpha)+\alpha \hat{v}_{w}^2}$, so the cold
beams with $\hat{v}_{w}\ll 1$ will reduce the growth rate of TEW since
$R<1$ for $\alpha>0$. Specifically, for $\alpha=\alpha_{max}\approx
1-2\hat{v}_c/\sqrt{\pi}$ with $\hat{v}_c\ll 1$, we have a reduced
factor of $R\approx \sqrt{2\hat{v}_c/\sqrt{\pi}+
  \hat{v}_{w}^2}$. Whereas, for $\hat{v}_c\gg 1$, the impact of cold
electron beams depends on $\alpha$. In the TQ, $\alpha$ is small with
$\alpha<\alpha_{max}\approx \exp(-\hat{v}_c^2)/(\sqrt{\pi}\hat{v}_c)$
for $\hat{v}_c\gg 1$. As such, the imaginary part of $D$, excluding
the factor $\omega_{pe}^2/\omega^2$, satisfies
 \begin{align}
\frac{\omega_r}{(kv_c)^2} \gamma -
\frac{kv_te^{-\hat{v}_c^2}}{2\sqrt{\pi} \gamma}
+\alpha\frac{\omega_{ce}}{2\gamma}=0.\label{eq-dispersion-trap-large-vc}
\end{align}
It follows that the infalling cold electrons of a tiny fraction in the
TQ will weaken the whistler instability through the third term in
Eq.~(\ref{eq-dispersion-trap-large-vc}). For a general finite $\alpha$
in other scenarios, the growth rate is determined by the real part of
$D,$
\begin{align}
\gamma=\left[1-(1-\alpha)\frac{\omega_r\omega_{pe}^2}{k^3v_cc^2}\right]^{-1/2}\frac{v_{w}}{2
  c} \omega_{pe},\label{eq-dispersion-trap-large-vc-finite-alpha}
\end{align}
which is the same as Eq.~(\ref{eq-growth-reflec-only-large-vc}) for
$\alpha= 1$. Since the factor in the bracket is larger than unity for
$\alpha<1$, the growth rate in
Eq.~(\ref{eq-dispersion-trap-large-vc-finite-alpha}) is smaller than
that in Eq.~(\ref{eq-growth-reflec-only-large-vc}) for $\alpha=1$ but
will be larger than that for $\alpha=0$.

\begin{figure}[h!]
\centering
\includegraphics[width=0.4\textwidth]{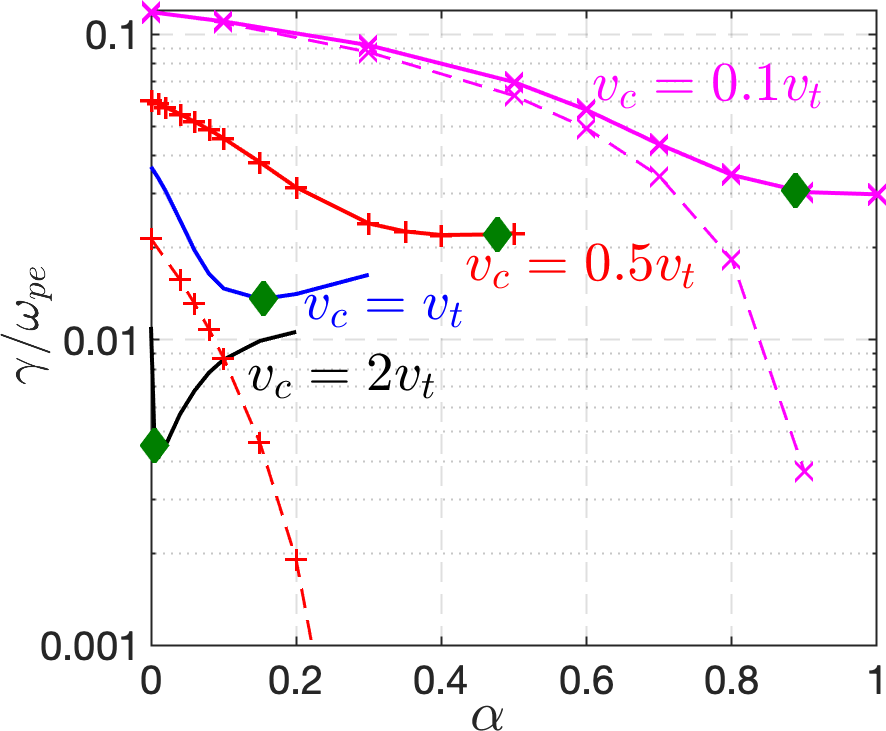}
\caption{Growth rates of the most unstable mode from
  Eq.~(\ref{eq-dispersion-trap-reflect}) are shown in solid lines for
  $\omega_{ce}=\omega_{pe}$, $c=5v_t$, $v_{w}=0.3v_t$ ($T_w\approx
  0.09 T_0$) and $\beta_0\equiv8\pi n_eT_0/B_0^2=4\%$, where the parameters correspond to the
  TQ simulations~\cite{Zhangfronts}. The diamonds label the upper
  limit of $\alpha=\alpha_{max}$ in the TQ process. For small $v_c
  =0.5 v_t$ and $0.1v_t$, growth rates for the {\em equivalent} TAW
  instability are shown in dashed lines. For $v_c\geq v_t$, {\em equivalent} TAW is
  stable and not shown.  }
\label{fig:growth_rate_vt_para-vc-1d5}
\end{figure}

Fig.~\ref{fig:growth_rate_vt_para-vc-1d5} shows the numerical
solutions of Eq.~(\ref{eq-dispersion-trap-reflect}) for $v_{w}=0.3v_t$
($T_w=0.09T_0$) but different $v_c$ and $\alpha$, which agree well
with
Eqs.~(\ref{eq-growth-reflect-trap-small-vc}-\ref{eq-dispersion-trap-large-vc-finite-alpha}). We
also plot the results from a bi-Maxwellian with equivalent
perpendicular and parallel temperatures, defined as
\begin{equation}
\frac{T_{e\perp}}{T_0}=R^2,~\frac{T_{e\parallel}}{T_0}=(1-\alpha)\left[1-\frac{2\hat{v}_c\exp(-\hat{v}_c^2)}{\sqrt{\pi}\textup{erf}(\hat{v}_c)}\right]
+ 2\alpha\hat{v}_c^2.\label{eq-equv-Tepara}
\end{equation}
It shows that the infalling cold electrons will also stabilize the {\em equivalent}
TAW instability, mainly
through the reduction of the temperature anisotropy for $\hat{v}_c>
\hat{v}_w*\sqrt{1/2-\hat{v}_c\exp(-\hat{v}_c^2)/(\sqrt{\pi}\textup{erf}(\hat{v}_c))}$,
which is readily satisfied for $\hat{v}_w\ll1$.  More importantly, it
shows that trapped electrons provide a more robust drive for the
whistler instability than temperature anisotropy even in the presence
of infalling cold electrons, so that the former will excite the
whistler waves first in a TQ. Another interesting and important
finding is that the growth rate of the most unstable mode will
increase with decreasing $v_c$ (due to the cooling of $T_{e\parallel}$
in the TQ) despite the infalling cold electrons (e.g., see the upper
bounds of $\alpha$ marked by the diamonds). Therefore, the whistler
instabilities and the associated wave-particle interactions will be
greatly enhanced with the cooling of $T_{e\parallel}.$

\begin{figure}[!htb]
  {\includegraphics[width=0.235\textwidth]{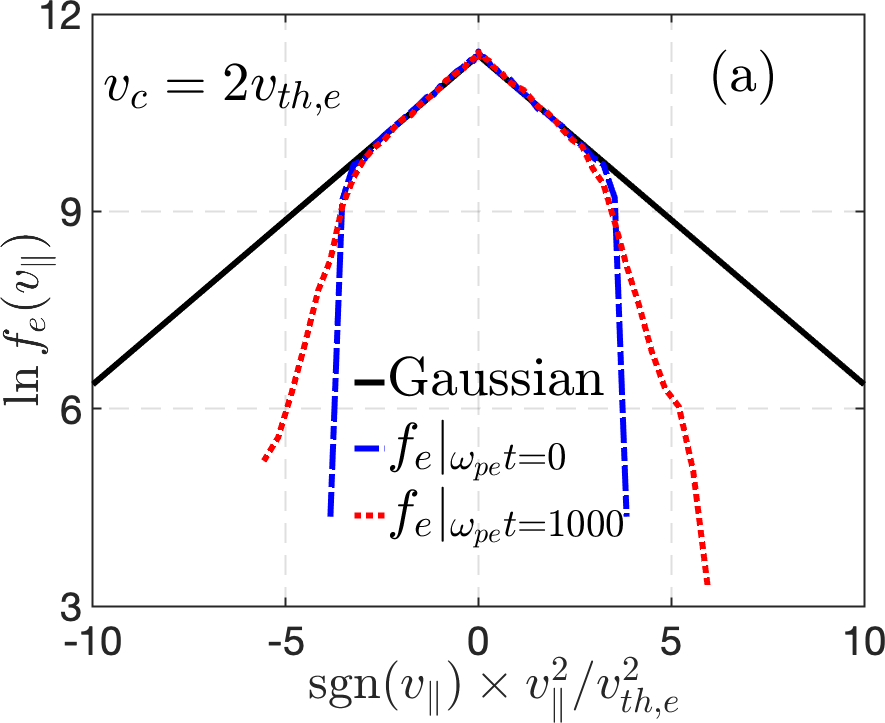}}
 {\includegraphics[width=0.235\textwidth]{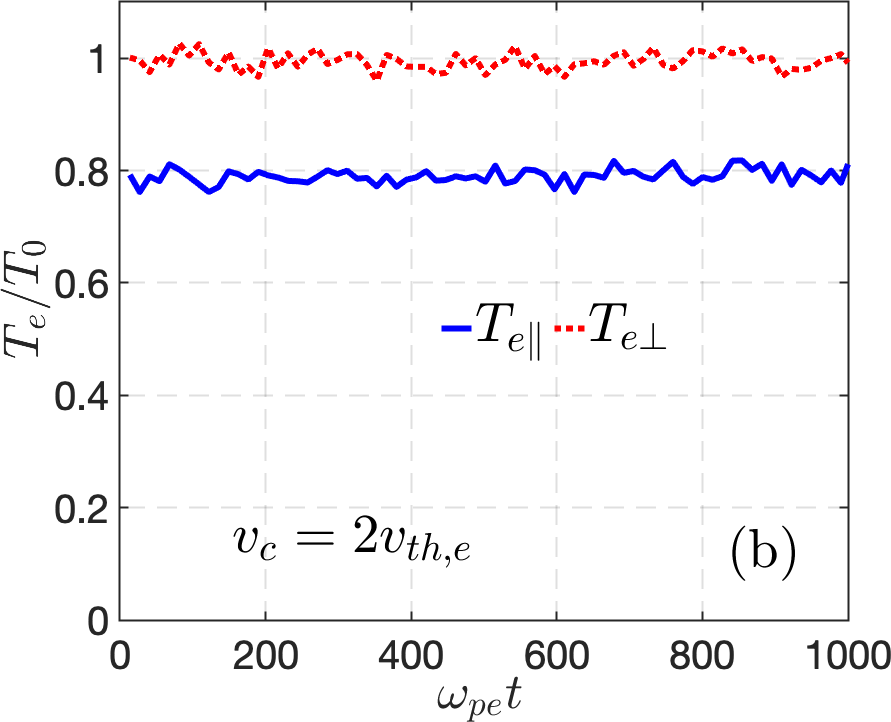}}\\
  {\includegraphics[width=0.235\textwidth]{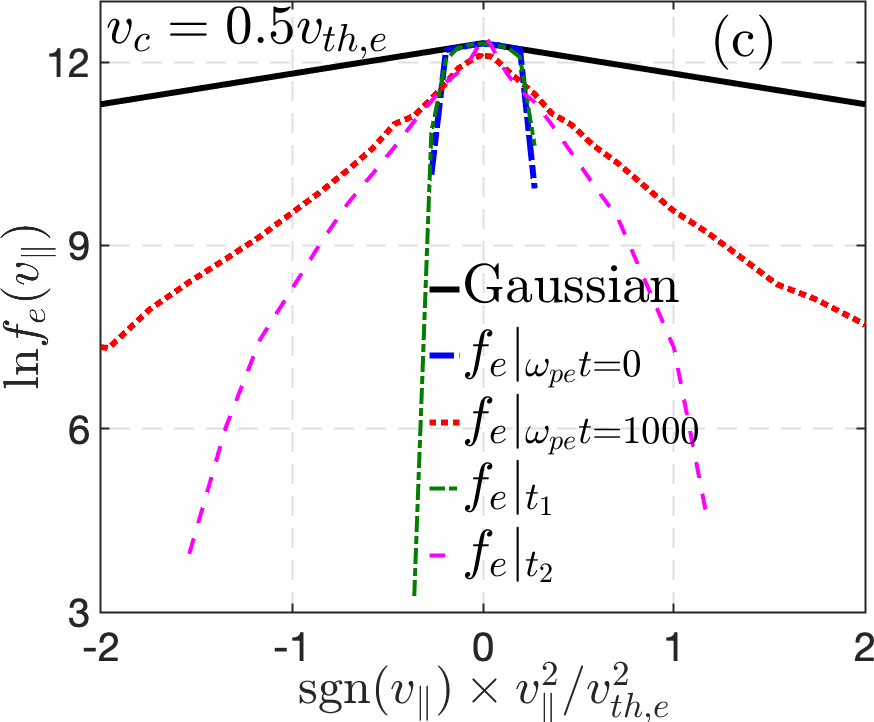}}
 {\includegraphics[width=0.235\textwidth]{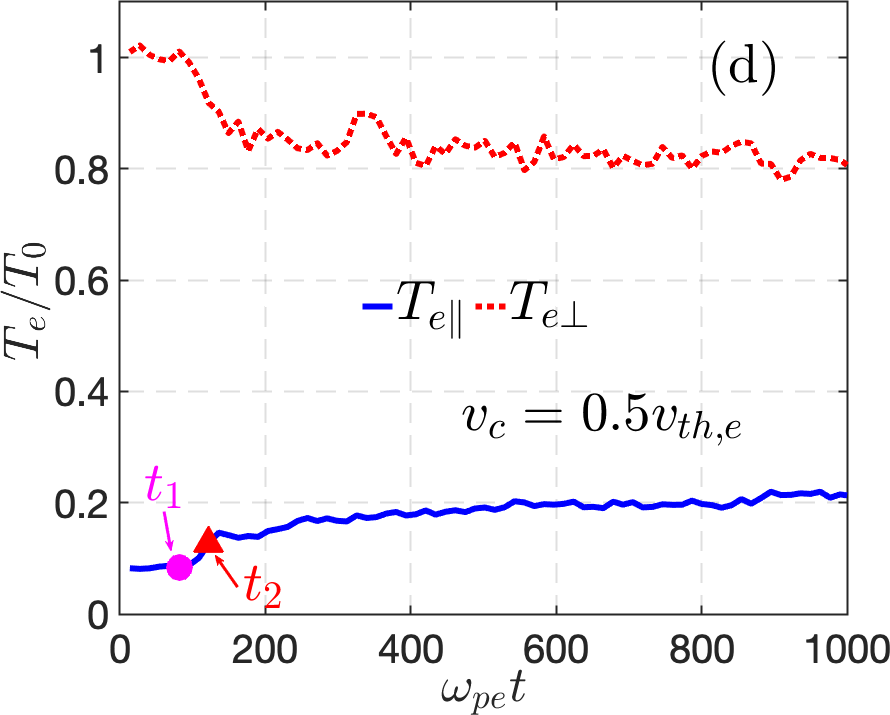}}
  \caption{$f_e(v_\parallel)$ and $T_{e\parallel,\perp}$ at plasma
    center are shown for the cases of $v_c=2v_{th,e}$ (a,b) and
    $0.5v_{th,e}$ (c,d) from periodic box simulations without infalling cold
    electrons (similar results exist for simulations with
    infalling cold electrons). For $v_c=0.5v_{th,e}$, $f_e$ and $T_e$ at $t_{1}=81\omega_{pe}^{-1}$ and $t_{2}=122\omega_{pe}^{-1}$ are shown. }
  \label{fig:whistler-in-periodic-box}
\end{figure}

{\em Two-stage process of $T_{e\perp}$ cooling by two kinds of
  whistler instabilities in sequence:} In the thermal quench of nearly
collisionless plasmas dominated by tail electron loss along the
magnetic field, the TEW instability, despite the stabilizing effect of
infalling cold electrons, will be excited first, for its much higher
growth rate. Interestingly, because the primary drive for this mode is
the sharp cut-off of the distribution at the electrostatic trapping boundary $v_\parallel=v_c$, it saturates
quickly with modest amount of smearing of the trapped-passing boundary~\cite{zhang-colli-damping}. Consequently there is
rather limited amount of $T_{e\perp}$ cooling if $v_c > v_{th,e}.$ To
illustrate this physics, we perform VPIC~\cite{VPIC} simulations in a periodic box
with initially truncated electron distribution given in
Eq.~(\ref{eq_cutoff_distribution}). Fig.~\ref{fig:whistler-in-periodic-box}(a,b)
shows the result for $v_c=2 v_{th,e},$ a case with unstable TEW mode
but no corresponding {\em equivalent} TAW instability (see
Fig.~\ref{fig:growth_rate_vt_para-vc-1d5}).  Nonlinear saturation of
the TEW modes produces a smeared cut-off boundary but no
appreciable temperature isotropization.

For deep cooling of $T_{e\parallel},$ which corresponds to $v_c$
significantly smaller than $v_{th,e},$ collisionless cooling of
$T_{e\perp}$ takes a two-stage route, which is shown in
Fig.~\ref{fig:whistler-in-periodic-box}(c,d) for the case of $v_c=0.5
v_{th,e}.$ The quick saturation of TEW from $t_1$ to $t_2$ produces an
approximate bi-Maxwellian with $T_{e\parallel} \ll T_{e\perp}$ by
$t_2.$ This is followed by the excitation of
TAW~\cite{kennel1966limit,gary1996whistler}, the saturation of which
produces further temperature isotropization after $t_2.$ The eventual
residual temperature anisotropy ($A_r\equiv T_{e\perp}/T_{e\parallel}$)
that this two-stage collisionless cooling of $T_{e\perp}$ can reach,
is set by the marginality condition for
TAW~\cite{gary1996whistler,gary2006linear}.

{\em Relative importance of different collisionless $T_{e\perp}$
  cooling mechanisms from fully kinetic TQ simulations:} In an
integrated TQ simulation over open magnetic field lines with connection
length $L_B,$ the dynamical cooling of both $T_{e\parallel}$ and
$T_{e\perp}$ is self-consistently accounted for. Since fully kinetic
VPIC~\cite{VPIC} simulations would need to resolve the Debye length
$\lambda_D,$ we will explore down-scaled simulations that retain the
extreme low collisionality of the physical system ($L_B/\lambda_{mfp} \ll 1$) but shrink the simulation
domain to $L_B/\lambda_D\sim 10^3.$
Previous comparison between theoretical analysis and simulation
results~\cite{Zhangfronts,listaged} has shown that such down-scaled
simulations capture the $T_{e\parallel}$ cooling physics accurately
for $L_B/\lambda_{mfp} \sim 0.05.$
Here we find that such
down-scaled simulations also allow the $T_{e\perp}$ cooling physics to
be reliably established, due to the large time-scale separation
between $T_{e\parallel}$ cooling and $(T_{e\perp},T_{e\parallel})$
isotropization by whistler instabilities. Recalling that during the
initial electron fronts phase~\cite{Zhangfronts} of the TQ, the rate
of $T_{e\parallel}$ cooling and hence $v_c$ reduction is $\sim
v_{th,e}/L_B$~\cite{listaged}. In reality, such a rate is much lower
than $\omega_{pe}$  due to the
large $L_B$. In contrast, the most unstable TEW/TAW mode has wavenumber
$k_m\sim \lambda_D^{-1}$ and growth rate $\gamma_m\sim
10^{-2}\omega_{pe}$ from our analysis.  The change in instability
drive ($v_c$) is thus much slower than the nonlinear saturation (that
occurs at $\sim 10^2 \omega_{pe}^{-1}$) of the whistler
instability. Such time-scale separation is even greater during the ion
fronts phase~\cite{Zhangfronts} of the TQ because the cooling of
$T_{e\parallel}$ is at an even lower rate, $\sim c_s/L_B$~\cite{listaged}. The down-scaled simulations, as
reported here, can resolve the most active whistler modes in space and
time over the slower process of $T_{e\parallel}$ cooling as long as
$L_B/\lambda_D \gtrsim 10^3$ is satisfied. An extra complication is
that the TEW mode is particularly sensitive to collisional
damping~\cite{zhang-colli-damping} so down-scaled simulations should
have $\lambda_{mfp} > 10^6\lambda_D,$ so $L_B/\lambda_{mfp}
< 10^{-3},$ which can be easily accommodated in down-scaled collisionless simulations.

\begin{figure*}[htb!]
\centering
{\includegraphics[width=0.31\textwidth]{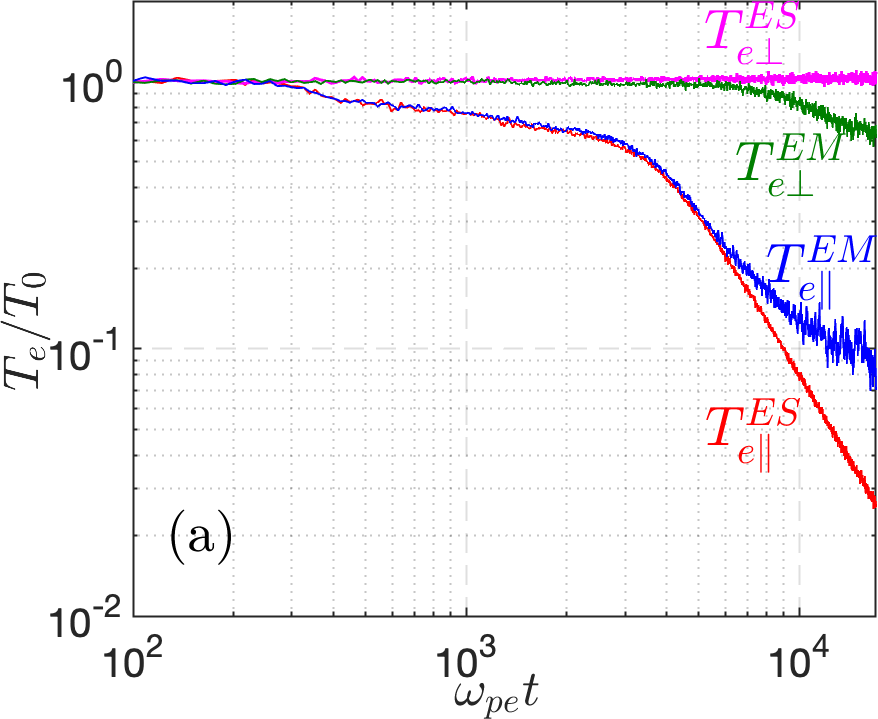}}
{\includegraphics[width=0.31\textwidth]{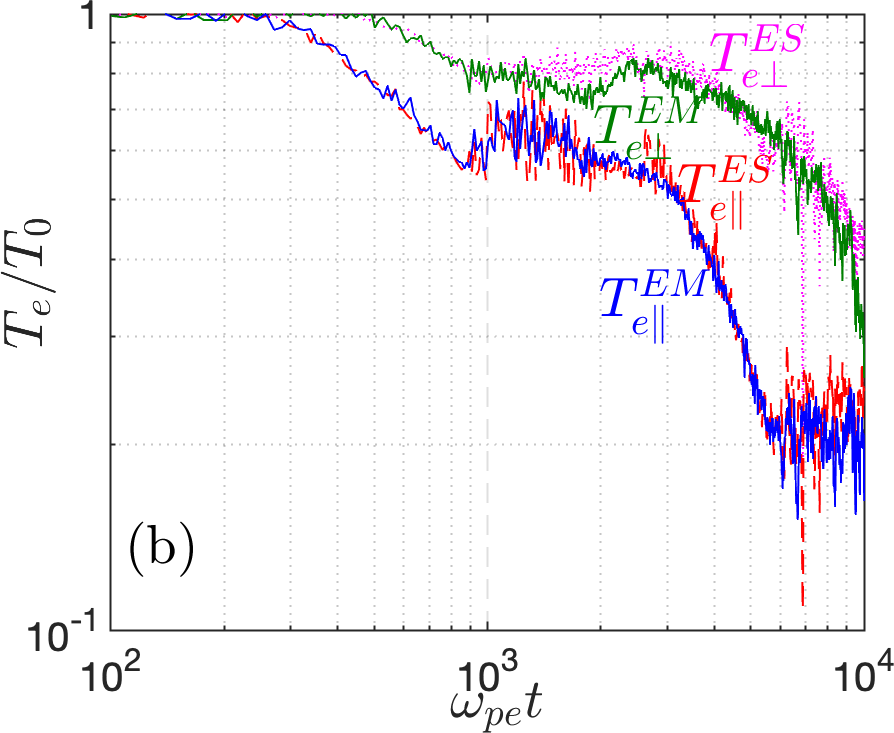}}
{\includegraphics[width=0.3\textwidth]{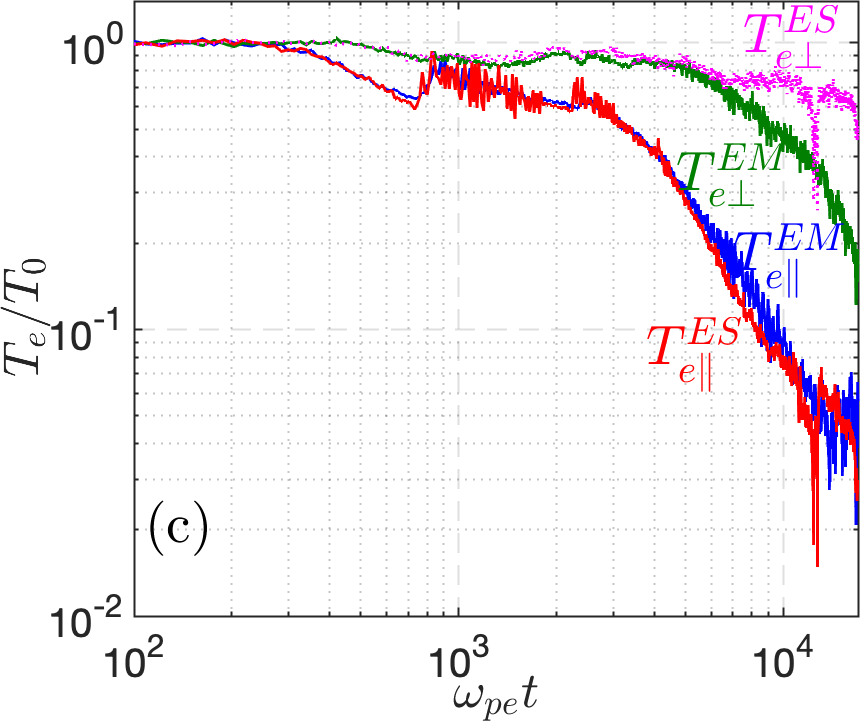}}
\caption{$T_{e\parallel,\perp}(t)$ at the center of plasmas from both electromagnetic
  (EM, with whistler modes) and electrostatic (ES, without whistler
  modes) TQ simulations. (a) is for the absorbing boundary, (b) and
  (c) are for plasma recycling boundary with $T_w=0.1T_0$ and
  $T_w=0.01T_0$, respectively. A reduced domain of
  $L_x=2L_B=1400\lambda_D$ and ion mass $m_i=100m_e$ are employed so
  that $\tau_{TQ}\equiv L_B/c_s \sim 10^4\omega_{pe}^{-1}$ for the ion
  front stage. Actual tokamak plasma of much longer $L_B$ has $\tau_{TQ}$ scaling up as $\propto L_B^{3/4}.$}
\label{fig:Te-center-absorbing}
\end{figure*}

To isolate the different cooling mechanisms in the integrated
$(T_{e\parallel}, T_{e\perp})$ cooling simulations, we contrast
electromagnetic (EM) with electrostatic (ES) simulations. Using a
plasma absorbing boundary to remove the possibility of dilutional
cooling, simulations in Fig.~\ref{fig:Te-center-absorbing}(a)
demonstrate that the whistler instabilities, retained in the EM
simulation but not the ES one, are self-excited to produce
collisionless $T_{e\perp}$ cooling.  With a plasma recycling boundary
condition to mimic a cooling boundary radiatively clamped to $T_w \ll
T_0$~\cite{Zhangfronts,listaged}, simulations in
Fig.~\ref{fig:Te-center-absorbing}(b,c) now retain the dilutional
cooling mechanism, which can dominate over that of the whistler
instabilities for modest amount of cooling, as seen in
Fig.~\ref{fig:Te-center-absorbing}(b) for $T_w/T_0=0.1.$
Fig.~\ref{fig:Te-center-absorbing}(c) reveals that dilutional cooling
(in the ES curve) is rather ineffective if deep cooling is needed,
which has $T_w/T_0=0.01$ in the simulation. The whistler instabilities
retained in the EM simulation, via the two-stage process noted earlier, are now
essential in the collisionless cooling of $T_{e\perp}$ that closely
tracks the cooling history of $T_{e\parallel}.$ 

\begin{figure}[h!]
\centering
\includegraphics[width=0.4\textwidth]{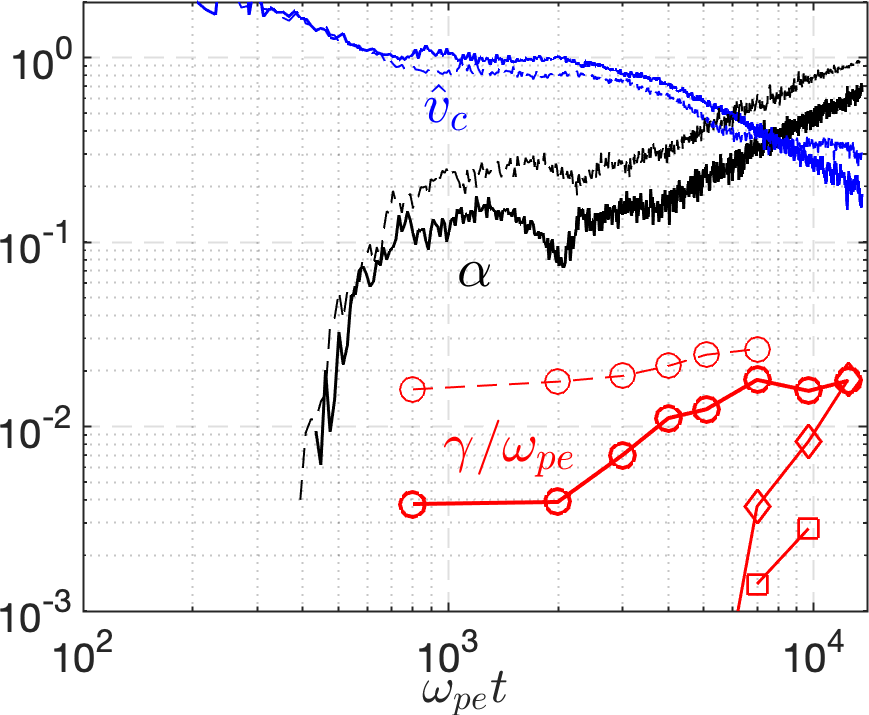}
\caption{Infalling cold 
  electron fraction $\alpha$ and the cutoff velocity (computed from
  Eq.~(\ref{eq-equv-Tepara})) at the plasma center from the EM simulation in Fig.~\ref{fig:Te-center-absorbing}b for
  $T_w=0.1T_0$ (dashed lines) and Fig.~\ref{fig:Te-center-absorbing}c
  for $T_w=0.01T_0$ (solid lines). From these $\alpha$ and $\hat{v}_c$,
  growth rates of the most unstable TEW and {\em equivalent} TAW modes
are computed (circles and diamonds, respectively).
{\em Equivalent} TAW for $T_w=0.1T_0$ is stable and thus not shown.
The growth rates of actual TAW using $T_{e\perp}$ from the $T_w=0.01T_0$ simulation
(squares) are much lower at late time because marginality is being approached.}
\label{fig:TQ-alpha-vc-growth}
\end{figure}

Fig.~\ref{fig:TQ-alpha-vc-growth} brings together the key
points that were introduced earlier in isolation.  Deep cooling of
$T_{e\parallel}$ comes from the drop of $v_c/v_t$ to $\sim 0.1,$
during which the infalling cold electrons take on an increasingly
larger fraction $\alpha \rightarrow 1.$ While the TEW instability is
robustly unstable from an early time, deep cooling of $T_{e\perp}$
requires the excitation of strong TAW modes in a two-stage
process. The residual $A_r$ is set by the
marginality of TAW. Specifically the marginal $A_r$ is a
decreasing function~\cite{gary1996whistler,gary2006linear} of
$\beta_{e\parallel}=8\pi n_eT_{e\parallel}/B_0^2$, i.e., $A_r=1+
S_e/\beta_{e\parallel}^{\zeta}$ with $\zeta \sim 0.5 $ and
$S_e\sim 0.5$ for $0.1\lesssim\beta_{e\parallel}\lesssim 1$.  For
smaller $\beta_{e\parallel}\sim 1\%$ in magnetic fusion plasmas, which
is further decreasing in the TQ process, the parameter $\zeta$ is
even smaller and VPIC simulations suggest that $\zeta\sim 0.2$. For
such small $\beta_{e\parallel}$, the energy conservation of plasma
indicates that $2T_{e\perp}+T_{e\parallel} \equiv T_{e\perp}(2+1/A_r)
= const.$. Recalling $A_r\gg 1$, $T_{e\perp}$ is slightly changed with a
fraction of $\sim 1/A_r$ but $T_{e\parallel}$ varies significantly due
to the temperature isotropization. Therefore, the more natural way to
express the bound of the temperature anisotropy should utilize
$\beta_{e\perp}\equiv A_r \beta_{e\parallel}$ rather than
$\beta_{e\parallel}$. As such, we have $A_r\approx
S_e'/\beta_{e\perp}^{\zeta'}$ with $\zeta'\sim 0.27$. The small
value of $\zeta'$ indicates that the saturated temperature
anisotropy will increase but remain a reasonably small value during
the TQ from $10$ KeV to $\sim 10s$ eV.

To conclude, we note that thermal quench experiments of Laboratory
magnetic fusion plasmas offers a rare opportunity to study the kinetic
transport physics underlying the conductive cooling of nearly
collisionless plasmas more commonly found in space and astrophysical
settings. There is a compelling case, from theory and simulation, that
collisionless cooling will bring down $T_{e\perp}$ proportionally to
follow a crashing $T_{e\parallel}.$ The specific mechanisms involve
dilutional cooling by infalling cold electrons, and wave-particle
interaction through a two-stage process driven by two kinds of
whistler wave instabilities.

We thank the U.S. Department of Energy Office of Fusion Energy
Sciences and Office of Advanced Scientific Computing Research
for support under the Tokamak Disruption Simulation (TDS) Scientific
Discovery through Advanced Computing (SciDAC) project, and the Base
Theory Program, both at Los Alamos National Laboratory (LANL) under
contract No. 89233218CNA000001. This research used
resources of the National Energy Research Scientific Computing Center
(NERSC), a U.S. Department of Energy Office of Science User Facility
operated under Contract No. DE-AC02-05CH11231 and the Los Alamos
National Laboratory Institutional Computing Program, which is
supported by the U.S. Department of Energy National Nuclear Security
Administration under Contract No. 89233218CNA000001.

\bibliography{reference,../cooling_front/reference}

\end{document}